# From Schrödinger's Equation to the Quantum Search Algorithm[1] [2]

Lov K. Grover *(lkgrover@bell-labs.com)*

**Abstract**

The quantum search algorithm is a technique for searching $N$ possibilities in only $O(\sqrt{N})$ steps. Although the algorithm itself is widely known, not so well known is the series of steps that first led to it, these are quite different from any of the generally known forms of the algorithm. This paper describes these steps, which start by discretizing Schrödinger's equation. This paper also provides a self-contained introduction to quantum computing algorithms from a new perspective.

**1. Introduction**  Consider the following problem from a crossword puzzle:

$$\_ \; \_ \; \mathbf{r} \; \_ \; \mathbf{n} \; \mathbf{h} \; \_ \qquad \text{(Solution - piranha)}$$

You have an online dictionary with 1, 000, 000 words in which the words are arranged alphabetically. You could program it to look for the solution to the puzzle so that it typically solves it after looking through 500, 000 words. It is very difficult to do much better than this. But that is: only if you limit yourself to a classical computer. A quantum computer can be in multiple states at the same time and, by proper design, can carry out multiple computations simultaneously. In case the above dictionary were available on a quantum computer, it would be possible to carry out the search in only about 1, 000 steps by using the quantum search algorithm.

The quantum search algorithm has evoked considerable interest among both physicists and computer scientists. This was the first important application of quantum computing that did not require the problem under consideration to have a structure and was hence applicable to several different types of problems in both physics and computer science. Also the framework was simple and general and could be extended to different problems and different physical situations.

It is unusual to write a paper listing the steps that led to a result after the result itself is well known. This is usually of historical interest and better left to the philosophers of science. In this case, more than five years after the initial algorithm was invented, the series of steps that led to it are still not known even by the scientists in the field. I recently presented an outline of this story to a small gathering of theoretical physicists[1]. After seeing the interest it inspired, I decided to write this more complete version.

**1(i). Quantum Mechanics**  The quantum search algorithm needs only a small fraction of the conceptual machinery of quantum mechanics. This subsection briefly mentions the concepts needed to understand the quantum search algorithm - it is by no means a comprehensive review of quantum mechanics.

---

1. A preliminary version of this paper was presented in the Winter Institute on the Foundations of Quantum Theory, SN Bose Center, Calcutta, India in Jan. 2000.
2. This material is based upon work supported in part by the NSA and ARO under contract no. DAAG55-98-C-0040.

A classical binary switch can be either ON or OFF. The two possibilities, ON / OFF, are referred to as *basis states* or sometimes just as *states* for short. The switch can be completely described by specifying which basis state it is in - whether it is ON or OFF. On the other hand a quantum mechanical system is associated with all possible basis states at the same time with certain probabilities - it can be simultaneously ON and OFF. Its specification requires us to specify the probabilities in all basis states. Further, as the system evolves, the probabilities in the various basis states interact with each other in complicated ways. This already gives some feeling of the complexity, and potential computational power, of quantum mechanical systems.

In order to harness the power of quantum mechanics, it is necessary to know how the probabilities in the various states interact with each other. The way to describe these probabilities is in "wave" terms, by a quantity called the amplitude. The amplitude is a complex number, with both a magnitude and a phase in each state. The specification of the amplitudes in each state is called a *superposition* or a *state vector*. The overall probability in each state, just like the intensity of a wave, is given by the absolute square of the amplitudes in that state. For example, consider a four state system. Denote the basis states of the system by $|0\rangle, |1\rangle, |2\rangle$ and $|3\rangle$. Let the amplitudes in these states be $\frac{1}{2}, \frac{1}{2}, \frac{1}{2}$ and $-\frac{1}{2}$ respectively. The probability in each of the four states is $\frac{1}{4}$, the state vector is $\begin{bmatrix} \frac{1}{2} \\ \frac{1}{2} \\ \frac{1}{2} \\ -\frac{1}{2} \end{bmatrix}$ which is also denoted as $\frac{1}{2}|0\rangle + \frac{1}{2}|1\rangle + \frac{1}{2}|2\rangle - \frac{1}{2}|3\rangle$. Note that the probabilities in each state would stay the same even if the state vector was $\frac{1}{2}|0\rangle + \frac{1}{2}|1\rangle + \frac{1}{2}|2\rangle + \frac{1}{2}|3\rangle$, however the state vector is very different and the evolution of the state vector in time will now be very different. In fact, the whole reason for describing the system in terms of amplitudes is that the evolution is better described in terms of amplitudes as opposed to probabilities.

Any physical process, whether classical or quantum mechanical, must evolve in time in a way so that it conserves the total probability. In Markov processes this requirement leads to the condition that the entries in each column of the state transition matrix sum to one. For quantum mechanical processes, this requirement translates into the condition that the state transition matrix be *unitary*, i.e. the columns of the state transition matrix be orthonormal. An example of a $4 \times 4$ unitary state transformation matrix is: $\frac{1}{2}\begin{bmatrix} 1 & 1 & 1 & 1 \\ 1 & 1 & -1 & -1 \\ 1 & -1 & -1 & 1 \\ 1 & -1 & 1 & -1 \end{bmatrix}$. This transforms $\frac{1}{2}|0\rangle + \frac{1}{2}|1\rangle + \frac{1}{2}|2\rangle + \frac{1}{2}|3\rangle$ into $|0\rangle$ and it transforms $\frac{1}{2}|0\rangle + \frac{1}{2}|1\rangle + \frac{1}{2}|2\rangle - \frac{1}{2}|3\rangle$ into $\frac{1}{2}|0\rangle + \frac{1}{2}|1\rangle - \frac{1}{2}|2\rangle + \frac{1}{2}|3\rangle$. Even though the initial superposi-

tions had the same probabilities in each state, after the quantum mechanical transformation the probabilities in the two cases become very different.

The transformation of amplitudes by any unitary operation is always linear. It is hence enough to specify how the basis states are transformed. The transformation of any superposition can be obtained from this simply by summing the transformations of each of the components, i.e. if $M$ be any operator, $|A\rangle$ and $|B\rangle$ be any two state vectors, then: $M(\alpha|A\rangle + \beta|B\rangle) = \alpha M|A\rangle + \beta M|B\rangle$. This is called the *superposition principle* and is the reason that in Dirac notation a superposition is denoted as an appropriately weighted summation of the components.

**1(ii). Quantum Computing** Just as classical computing systems are synthesized out of two-state systems called bits, quantum computing systems are synthesized out of two-state systems called *qubits*. The difference is that a bit can be in only one of the two states at a time, on the other hand a qubit can be in both states at the same time. For example, consider two qubits, each in the superposition: $\left(\frac{1}{\sqrt{2}}|0\rangle + \frac{1}{\sqrt{2}}|1\rangle\right)$. This is represented as the tensor product: $\left(\frac{1}{\sqrt{2}}|0\rangle + \frac{1}{\sqrt{2}}|1\rangle\right) \otimes \left(\frac{1}{\sqrt{2}}|0\rangle + \frac{1}{\sqrt{2}}|1\rangle\right)$ or equivalently we have a four-state system in a superposition: $\frac{1}{2}|0\rangle \otimes |0\rangle + \frac{1}{2}|0\rangle \otimes |1\rangle + \frac{1}{2}|1\rangle \otimes |0\rangle + \frac{1}{2}|1\rangle \otimes |1\rangle$ which is also represented as $\frac{1}{2}|00\rangle + \frac{1}{2}|01\rangle + \frac{1}{2}|10\rangle + \frac{1}{2}|11\rangle$. This is an example of a superposition that *can* be factored into smaller systems. However, most superpositions, such as $\frac{1}{2}|00\rangle + \frac{1}{2}|01\rangle - \frac{1}{2}|10\rangle + \frac{1}{2}|11\rangle$, *cannot* be represented as a product of smaller superpositions. The general description of an $n$ qubit system (which is a $2^n$ state system) requires us to specify the amplitude in each of $2^n$ states, i.e. $2^n$ quantities; whereas to specify a classical $n$ bit system requires just $n$ bits of information.

Interesting and paradoxical quantum mechanical effects arise in systems that cannot be factored into smaller systems. Such systems are called entangled. Quantum computation algorithms, such as quantum search, make use of entanglement to devise fast algorithms.

In order to design quantum computing systems, we need a basic set of building blocks analogous to the NAND and NOR gates that are used to build classical digital systems. Unfortunately, by writing out the transformation matrices for NAND and NOR, it is easily seen that they are not unitary and hence cannot be implemented quantum mechanically. Fortunately, there exists a set of unitary transformations that can be implemented quantum mechanically, using which it is possible to design a circuit that will output any desired boolean function. This is the following set of three transformations:

(i) NOT - a one-input one-output gate. Since the input and output are both qubits, this gate transforms a two state system into another two state system. The transformation swaps $|0\rangle$ and $|1\rangle$. In matrix terms this is described by the transformation $\begin{bmatrix} 0 & 1 \\ 1 & 0 \end{bmatrix}$ where the basis states are $|0\rangle$ and $|1\rangle$ respectively.

(ii) CNTRL-NOT - a two-input two-output gate: the input and output both consist of two qubits - hence this is a

transformation from one four-state system to another four-state system. It transforms $|00\rangle$ into $|00\rangle$; $|01\rangle$ into $|01\rangle$; $|10\rangle$ into $|11\rangle$; $|11\rangle$ into $|10\rangle$. The transformation matrix for this is $\begin{bmatrix} 1 & 0 & 0 & 0 \\ 0 & 1 & 0 & 0 \\ 0 & 0 & 0 & 1 \\ 0 & 0 & 1 & 0 \end{bmatrix}$, where the basis states are $|00\rangle$, $|01\rangle$, $|10\rangle$ and $|11\rangle$ respectively. The name CNTRL-NOT arises since the first bit acts as a *control* for the second - if it is a 1, it swaps 0 and 1 in the second bit. It must be emphasized that this simple-minded verbal description only holds for the basis vectors. The transformations of superpositions are more complicated and can only be obtained by the state transition matrix described above.

(iii) CNTRL-CNTRL-NOT - a three-input three-output gate: the input and output both consist of three qubits - hence this transforms one eight-state system to another. It transforms $|000\rangle$ into $|000\rangle$; $|001\rangle$ into $|001\rangle$; $|010\rangle$ into $|010\rangle$; $|011\rangle$ into $|011\rangle$; $|100\rangle$ into $|100\rangle$; $|101\rangle$ into $|101\rangle$; $|110\rangle$ into $|111\rangle$; $|111\rangle$ into $|110\rangle$. Just like (i) and (ii), this too can be listed out as an $8 \times 8$ matrix transformation (an $8 \times 8$ identity matrix with the last two columns swapped). The name CNTRL-CNTRL-NOT arises since the first two bits act as *control bits* for the third - if both are 1, it swaps 0 and 1 in the third bit. As mentioned before, this simple verbal description only holds for the basis vectors - for superpositions, we need to use the state transition matrix.

The unitarity of these three operations is easily verified by noting that the column vectors of the transformation matrices are orthonormal (this is clearly the case since the gates merely permute the basis states). Using these three gates it is possible to design a circuit such that one of its outputs is any specified boolean function $f(\bar{x})$. This needs approximately the same number of gates as would be required in a classical implementation using NANDs and NORs. Such a circuit maps a superposition that is concentrated in any single input basis state, $\bar{x}$, to another superposition concentrated in a single output basis state $f(\bar{x})$. Transformations of general superpositions are obtained from the superposition principle. Note that to synthesize boolean functions quantum mechanically, we need three basic gates whereas in the classical case we needed just two (NAND and NOR).

In order to develop more powerful quantum mechanical algorithms, in addition to these three gates, we need some operations that are essentially quantum mechanical and have no classical analog, i.e. the entries of the state transition matrix are not all 0's and 1's. Two such operations that we need in the quantum search algorithm are the W-H transformation operation and the selective inversion operation, these are discussed in the following paragraphs.

A basic operation in quantum computing is the operation $M$ performed on a single qubit - this is represented by the following matrix: $M \equiv \frac{1}{\sqrt{2}} \begin{bmatrix} 1 & 1 \\ 1 & -1 \end{bmatrix}$. $M$ is unitary since the columns are orthonormal. Also note that $MM = I$. If we consider an $n$ qubit system, we can perform the transformation $M$ on each qubit independently in sequence, thus transforming the state of the system. A system consisting of $n$ qubits has $N \equiv 2^n$ basis states, so the state transition matrix representing this operation is of dimension $2^n \times 2^n$. Consider the case when the starting state is one of the $2^n$ basis states, i.e. it is described by an arbitrary string of $n$ binary digits composed of some 0s and some 1s. The result of performing the transformation $M$ on each qubit will be a superposition of states consisting of all possible $n$ bit binary strings with amplitude of each state being $\pm 2^{-\frac{n}{2}}$. This transformation is referred to as the W-H transformation and denoted by $W$. Note that since $MM = I$, it follows that $WW = I$. Also note that, since $M = M^T$, it follows that $W = W^T$. A generalization of this is the Quantum Fourier Transformation, which leads to applications such as factorization.

The other transformation that we need is the selective phase inversion of the amplitude in certain marked states. The transformation matrix describing this for a four state system with selective phase inversion of the second state is: $\begin{bmatrix} 1 & 0 & 0 & 0 \\ 0 & -1 & 0 & 0 \\ 0 & 0 & 1 & 0 \\ 0 & 0 & 0 & 1 \end{bmatrix}$. This is clearly unitary. Unlike the W-H transformation, in this case, the probability in each state stays the same.

Consider a binary function $f(\bar{x})$ that is either 0 or 1. A quantum mechanical circuit can be designed to invert the amplitude in the set of basis states, $\bar{x}$, for which $f(\bar{x}) = 1$ - provided we are given a quantum mechanical black box that evaluates the function $f(\bar{x})$ for any specified $\bar{x}$ such as that in Figure 1. Note that we do not need to know in advance which $\bar{x}$ makes the function equal to 1 - all we need is a circuit which for an arbitrary $\bar{x}$, can tell us whether or not $f(\bar{x})$ is 1. Given such a quantum mechanical black box, a selective inversion transformation can be realized by using this box along with some of the gates discussed so far - this is described in the circuit in Figure 5 (in appendix).

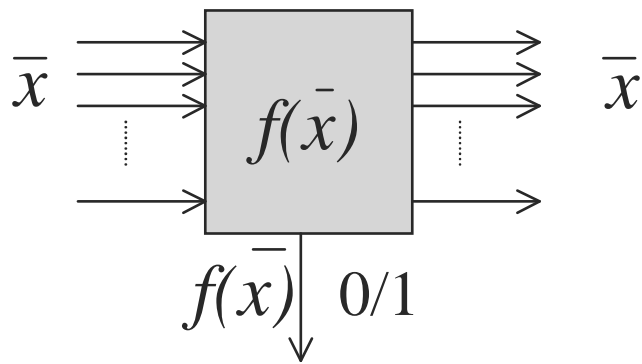

*Figure 1 - A quantum mechanical "black box" that evaluates $f(\bar{x})$.*

**1(iii) Quantum Search** As mentioned in the introduction, in the quantum search algorithm we are given a certain condition and we need to find out which one of the specified $N$ possibilities satisfies this. This problem can be represented as a binary function $f(\bar{x})$ defined over $N$ basis states (denoted by $\bar{x}$), $f(\bar{x})$ is known to be 1 at a single value of $\bar{x}$, say $t$ ($t$ for target) - the goal is to find $t$. Without any other information about the structure of $f(\bar{x})$, it would take an average of $\frac{N}{2}$ function evaluations to solve this problem on a classical computer. [1] found a quantum mechanical algorithm that took only $O(\sqrt{N})$ steps.

This was of considerable interest since, by properly defining the function $f(\bar{x})$, a number of different problems, such as the crossword puzzle mentioned in the introduction, could be cast in this form. All that is needed is a quantum mechanical box for evaluating the function $f(\bar{x})$. By putting a small amount of hardware around this box it is possible to search $N$ possibilities in only $O(\sqrt{N})$ steps. The only information about $f(\bar{x})$ needed is that it is a binary function that is either 0 or 1 and that it is 1 at only a single point in the domain.

The quantum search algorithm consisted of $\sqrt{N}$ repetitions of the operator $-I_{\bar{0}}WI_fW$ starting with the basis state where all qubits are 0 - this basis state is denoted by $\bar{0}$. $W$ denotes the Walsh-Hadamard (W-H) transformation, $I_f$ denotes the selective phase inversion of the target state $t$ where the function $f(\bar{x})$ evaluates to 1, $I_{\bar{0}}$ denotes the selective phase inversion of $\bar{0}$. This creates a superposition all of whose amplitude is in the basis state $t$. A measurement then immediately identifies $t$.

It is easily possible to carry out the algebra and verify that the evolution of the system is such that in every repetition of $-I_{\bar{0}}WI_fW$, the amplitude in the $t$ state rises by $O\left(\frac{1}{\sqrt{N}}\right)$ [1]. Unfortunately, this calculation does not give much insight into how such an algorithm could have been first invented. The algorithm was first presented in a conference in terms of a diffusion transform [2], which is similar to the way it was invented, but the significance of this was not usually appreciated. Later on, there were more interpretations: inversion about average, rotation in a two dimensional Hilbert space, antenna array [3]. All of these describe some aspect of the algorithm, but they are very different from the way it was initially invented.

**2. Schrödinger's Equation** Quantum mechanics was historically developed in the context of atomic physics. One of the first successful descriptions of atomic phenomena was the Schrödinger's Equation which was discovered by Erwin Schrödinger in 1926 and to this day it continues to be the most commonly used description for non-relativistic phenomena. This was the aspect of quantum mechanics that I was familiar with when I started investigating quantum algorithms in 1995.

In Schrödinger's framework, the basis states are continuous and uniformly distributed in space. Instead of the discrete states $|0\rangle$ and $|1\rangle$, there is a continuum of states denoted by $|x\rangle$ where $x$ is a continuous variable. The state vector is specified in the form of a function $\psi(x)$ which denotes the amplitude in the state $|x\rangle$ (this function is usually referred to as a wavefunction). Schrödinger's Equation describes the evolution of the wavefunction $\psi(x)$ in time for actual physical systems which are described by potential functions.

Let us begin with Schrödinger's Equation in one dimension (leaving out the scaling constants in order to focus on the nature of the equation): $i\frac{\partial}{\partial t}\psi(x, t) = -\frac{\partial^2}{\partial x^2}\psi(x, t) + V(x)\psi(x, t)$, which may be written as $\frac{\partial}{\partial t}\psi(x, t) = i\frac{\partial^2}{\partial x^2}\psi(x, t) - iV(x)\psi(x, t)$. This describes the evolution of the wavefunction $\psi(x, t)$ in the presence of a potential function: $V(x)$.

This is analogous to the diffusion equation with absorption. If we break up the evolution into infinitesimal time steps of size $dt$ and imagine that each point is connected to its two adjacent neighbors on an infinitesimal grid of size $dx$, then the derivatives in Schrödinger's equation can be written as: $\frac{\partial}{\partial t}\psi(x, t) \equiv \frac{\psi(x, t + dt) - \psi(x, t)}{dt}$ and $\frac{\partial^2}{\partial x^2}\psi(x, t) \equiv \frac{\psi(x + dx, t) + \psi(x - dx, t) - 2\psi(x, t)}{(dx)^2}$. It simplifies notation if we define a constant $\varepsilon$ such that $dx \equiv \sqrt{\frac{dt}{\varepsilon}}$. Substituting the definitions for the derivatives into Schrödinger's equation, we obtain the evolution of the wave function with time, i.e. $\psi(x, t + dt) = (1 - iV(x)dt - 2i\varepsilon)\psi(x, t) + i\varepsilon(\psi(x + dx, t) + \psi(x - dx, t))$.

In the Schrödinger's equation there is a continuum and therefore an infinite number of states; however in order to demonstrate the principle, assume just a 4-state system where the states are arranged in a loop with diffusion from each state to its two adjacent states. $[\psi(x, t)]$ is a column vector with 4 components. Then in accordance with Schrödinger's equation the evolution for a time $dt$ may be represented as a state matrix transformation as follows:

$$[\psi(x, t+dt)] = \begin{bmatrix} 1 - iV(x_1)dt - 2i\varepsilon & i\varepsilon & 0 & i\varepsilon \\ i\varepsilon & 1 - iV(x_2)dt - 2i\varepsilon & i\varepsilon & 0 \\ 0 & i\varepsilon & 1 - iV(x_3)dt - 2i\varepsilon & i\varepsilon \\ i\varepsilon & 0 & i\varepsilon & 1 - iV(x_4)dt - 2i\varepsilon \end{bmatrix} [\psi(x, t)]$$

It may be verified that this is a unitary transformation in the limit of infinitesimal $dt$ and infinitesimal $\varepsilon$, i.e. the magnitude of each column vector is $1 + O((dt)^2) + O(\varepsilon^2)$ and the dot product of any two column vectors is $O((dt)^2) + O(\varepsilon^2)$.

The above state transition matrix may be represented approximately as follows:

$$[\psi(x, t+dt)] = DR[\psi(x, t)], \text{ where:}$$

$$D \equiv \begin{bmatrix} 1-2i\varepsilon & i\varepsilon & 0 & i\varepsilon \\ i\varepsilon & 1-2i\varepsilon & i\varepsilon & 0 \\ 0 & i\varepsilon & 1-2i\varepsilon & i\varepsilon \\ i\varepsilon & 0 & i\varepsilon & 1-2i\varepsilon \end{bmatrix}, R \equiv \begin{bmatrix} \exp(-iV(x_1)dt) & 0 & 0 & 0 \\ 0 & \exp(-iV(x_2)dt) & 0 & 0 \\ 0 & 0 & \exp(-iV(x_3)dt) & 0 \\ 0 & 0 & 0 & \exp(-iV(x_4)dt) \end{bmatrix}.$$

The representation is accurate up to terms $O((dt)^2) + O(\varepsilon^2)$. Also note that $R$ is exactly unitary, but $D$ only approximately so, up to $O(\varepsilon^2)$.

The sum of the entries in each column of $D$ is unity so it is like a Markov diffusion process with imaginary state transition probabilities ($D$ is short for *Diffusion Transform)*. This Markov diffusion feature will be made use of later in the design of the algorithm.

Repeating the infinitesimal state transformation, we obtain the transformation for finite times:

$$[\psi(x,\tau)] = (DR)......\frac{\tau}{dt} repetitions......(DR)(DR)(DR)[\psi(x,0)]$$

Note that $D$ and $R$ do not commute, which is why in the equation above they have to be carefully listed out in the indicated order. In theoretical physics this technique for building up finite transformations out of infinitesimal non-commuting operations is called Trotter's formula.

## 3. Infinitesimal Quantum Search
By analogy with a classical situation, we know that if we start with a uniform superposition in a line and let it evolve in the presence of a potential function, it will gravitate towards points at which the potential is lower (see Figure 2.)

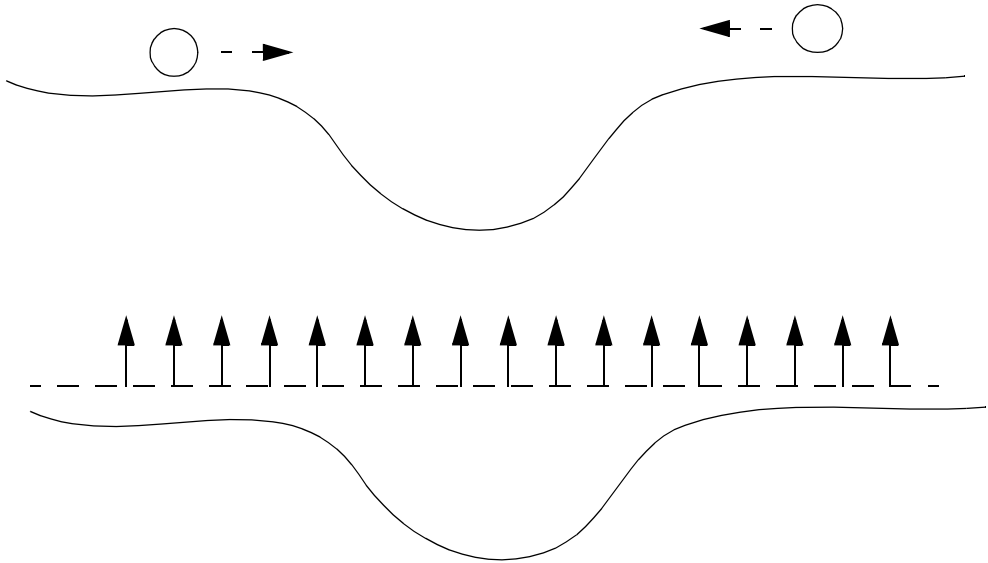

*Figure 2 - Just as the balls roll down into regions of lower potential energy, a uniform quantum superposition evolving under Schrödinger's equation will gravitate towards lower potential energy regions*.

Therefore, in order to design an algorithm that will reach certain marked states, one can give these marked states a lower potential and then implement the same iterated transformations obtained from the evolution of Schrödinger's equation. We assume a single marked state. Assume we do not know which one this is, but we have a means of selectively rotating its phase by any desired amount (the implementation of such a transformation is discussed in section 4). Using such a phase rotation transformation, $R$, we design an algorithm using the analogy with Schrödinger's equation.

Consider the same sequence of infinitesimal transformations used to represent the evolution of a system in the presence of a potential (last equation of the previous section). In the design of a quantum algorithm an immediate improvement is obtained by noticing that we can connect each state to all other states, instead of just to states that are adjacent on the grid, thus synthesizing an algorithm of the type:

$$[\psi(\bar{x}, \tau)] = (DR)\ldots\ldots\ldots(DR)(DR)(DR)[\psi(\bar{x}, 0)]$$

where:[1] $D \equiv \begin{bmatrix} (1-iN\varepsilon) & i\varepsilon & i\varepsilon & \ldots & i\varepsilon \\ i\varepsilon & (1-iN\varepsilon) & i\varepsilon & \ldots & i\varepsilon \\ i\varepsilon & i\varepsilon & (1-iN\varepsilon) & \ldots & i\varepsilon \\ \ldots & \ldots & \ldots & \ldots & \ldots \\ i\varepsilon & i\varepsilon & i\varepsilon & \ldots & (1-iN\varepsilon) \end{bmatrix}$ and $R \equiv \begin{bmatrix} 1 & 0 & 0 & \ldots & 0 \\ 0 & \exp(i\gamma) & 0 & \ldots & 0 \\ 0 & 0 & 1 & \ldots & 0 \\ \ldots & \ldots & \ldots & \ldots & \ldots \\ 0 & 0 & 0 & \ldots & 1 \end{bmatrix}$

Note that according to the state-transition transformation derived from Schrödinger's Equation, positive values of $\gamma$ correspond to negative potentials. The question is regarding whether such a series of transformations will really drive the system into the marked state and if so how large $\varepsilon$ and $\gamma$ should be and how many times the $(DR)$ operation has to be repeated.

Assume $[\psi(\bar{x}, 0)]$ to be the uniform superposition, that is, a wavefunction with equal amplitudes in all states. The effect of $R$ is to rotate the phase of the target state by $\gamma$. Note that the sum of entries in each column of $D$ is unity, hence the operation $D$ is analogous to a Markov diffusion process with a transition probability from any state to any other state being $i\varepsilon$. This analogy simplifies the analysis and design of the algorithm. It immediately follows that there is no net exchange between states that have the same amplitude since the transfer in the two directions cancels out, the only net exchange is between the state with the rotated phase and each of the other states.

The increase of amplitude in the marked state due to $D$ is maximum when $R$ initially rotates its phase by $\frac{\pi}{2}$ - the increase of amplitude is then proportional to $\varepsilon$ as shown in Figure 3. Let the amplitudes in the unmarked and marked states be $\frac{k}{\sqrt{N}}$ and $\frac{iK}{\sqrt{N}}$ respectively. Initially $k$ and $K$ are both 1 but as the amplitude in the marked state increases, $K$ rises to $O(\sqrt{N})$ and $k$ shrinks.

---

1. For an $N$-state system, in order for the terms in each row to sum to unity, the diagonal terms of $D$ should be $(1-i(N-1)\varepsilon)$. However, for simplicity, we write this as $(1-iN\varepsilon)$. For $N$ large, the error due to this is negligible.

Summing the transfers between each of the unmarked states and the marked state, it follows that after $D$, the amplitude in the marked state becomes $\frac{iK}{\sqrt{N}} + \frac{ik}{\sqrt{N}}N\varepsilon + \frac{K}{\sqrt{N}}N\varepsilon$ and the amplitude in each of the other states becomes $\frac{k}{\sqrt{N}} - \frac{K}{\sqrt{N}}\varepsilon - \frac{ik}{\sqrt{N}}\varepsilon$. Assuming $k$ to be of order 1 (as is the case initially), it follows that there is a net change of magnitude of approximately $\frac{N\varepsilon}{\sqrt{N}}$ in the amplitude of the marked state. Also, the phase of the marked state gets rotated by approximately $N\varepsilon$, the magnitude and phase in the unmarked states stay approximately the same. ($k$, the magnitude of the amplitudes in the unmarked states, is easily estimated by a conservation of probability argument, i.e. the quantity $k^2 + \frac{K^2}{N}$ is the total probability and is always 1. Therefore, as long as $K < \sqrt{N/2}$, $k$ lies in the range $1/2 < k < 1$).

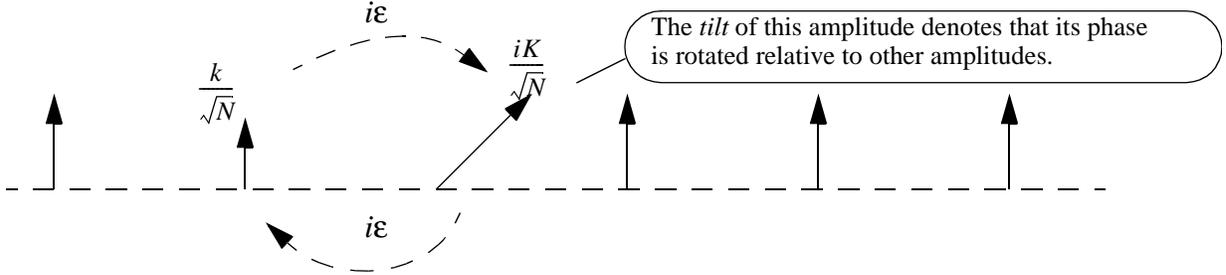

*Figure 3 - The diffusion operation (D) results in a net transfer of amplitude proportional to $\varepsilon$ from each state to the marked state whose phase has been rotated by R. There is no net transfer between states with equal amplitude.*

The previous paragraph describes how it is possible to obtain a transfer into the marked state even without knowing which one it is, provided we can somehow rotate the phase of the marked state and subsequently apply a diffusion operation, $D$, that transfers amplitude proportional to $i\varepsilon$ between any two states. The analysis of the previous paragraph suggests that the maximum transfer is achieved by keeping $\varepsilon$ as high as possible which might accomplish the entire transfer in a single operation. The problem, however, is that the matrix $D$, as defined in the previous section, is no longer unitary for large $\varepsilon$. If we look at the columns, their dot product is $O(N\varepsilon^2)$ and the sum of the squares of magnitudes in each column is $1 + O(N^2\varepsilon^2)$. $D$ is therefore only approximately unitary provided $\varepsilon \ll \frac{1}{N}$.

The transfer of amplitude into the state with the rotated phase was estimated as $\frac{N\varepsilon}{\sqrt{N}}$. Assuming $\varepsilon = O\left(\frac{1}{N}\right)$, this transfer becomes $O\left(\frac{1}{\sqrt{N}}\right)$. The rotation of the phase of the marked state which was $N\varepsilon$, becomes of order 1. If we adjust the $R$ matrix to *unrotate* the phase by this precise amount, then we can repeat the $D$ operation to obtain an additional transfer. It follows that in $O(\sqrt{N})$ repetitions of the $(DR)$ cycle, the amplitude and thus probability in the marked state will become a quantity of order 1.

## 4. An Exactly Unitary Diffusion Transformation

In the previous section, there is the tricky matter of precisely how high $\varepsilon$ can be. The issue is regarding the fact that the matrix $D$ is unitary only up to $O(N^2\varepsilon^2)$ and therefore $\varepsilon$ is required to be much smaller than $\frac{1}{N}$. The number of steps required by the algorithm is $O\left(\frac{\sqrt{N}}{N\varepsilon}\right)$ which can therefore at best be a large constant times $\sqrt{N}$ - the precise scaling factor will depend on precisely how much $\varepsilon$ is smaller than $\frac{1}{N}$.

We could carry out a perturbation to make $D$ unitary to higher powers of $\varepsilon$ thus making it possible to make $\varepsilon$ higher. Instead, this section will *create* a diffusion matrix that is exactly unitary.

Consider $D \equiv \begin{bmatrix} a & b & b & \ldots & b \\ b & a & b & \ldots & b \\ b & b & a & \ldots & b \\ \ldots & \ldots & \ldots & \ldots & \ldots \\ b & b & b & \ldots & a \end{bmatrix}$ where $a$ and $b$ are complex constants to be determined based on the unitarity condition. Note that the structure of this matrix immediately implies that the sum of the entries in each column be $\exp(i\lambda)$ where $\lambda$ is real (if $\lambda$ had an imaginary component, it would imply that the sum of the amplitudes would increase or decrease exponentially as the wavefunction evolved). Therefore by performing a phase rotation transformation on $D$ we can always transform it to a form where the sum of the entries in each column add to 1.

The condition that the sum of the absolute squares of each column vector of $D$ is unity, gives equation (i) below and the condition that any two column vectors of $D$ be orthogonal yields equation (ii).

$|a|^2 + (N-1)|b|^2 = 1$ \qquad (i)

$2Real(ab^*) + (N-2)|b|^2 = 0$ \qquad (ii)

The case previously considered in section 3 ($a = 1 - iN\varepsilon$; $b = i\varepsilon$) satisfies these two conditions approximately provided $\varepsilon \ll \frac{1}{N}$. We would like to make $|b|$ as high as possible in order to attain the maximum transfer in each step. This section shows that $|b|$ can be made as high as $\frac{2}{N}$.

There is some freedom in the choice of $a$ and $b$ since there are 4 variables (the real and imaginary portions of $a$ and $b$) and only 2 equations. One of the degrees of freedom represents the fact that if we have any solution $(a, b)$ where $a$ and $b$ are complex, then $(a\exp(i\phi), b\exp(i\phi))$ is also a solution for any real $\phi$. Therefore, we can choose the phase of one of the variables arbitrarily, say we choose $a$ to be real. The equations (i) and (ii) then become:

$$a_r^2 + (N-1)(b_r^2 + b_i^2) = 1 \qquad \text{(i.a)}$$

$$2a_r b_r + (N-2)(b_r^2 + b_i^2) = 0 \qquad \text{(ii.a)}$$

Substituting for $a_r$ in terms of $|b|^2$ from (i.a) into (ii.a) and squaring both sides, we obtain:

$$\frac{(N-2)^2 |b|^4}{1-(N-1)|b|^2} = 4b_r^2$$

$$\leq 4|b|^2$$

This leads to the bound $|b| \leq \frac{2}{N}$ which implies $|b_r| \leq \frac{2}{N}$. Trying out $b_r = \frac{2}{N}$, $b_i = 0$ indeed gives a solution with $a = -1 + \frac{2}{N}$ and $D$ becomes:

$$D = \begin{bmatrix} \left(-1+\frac{2}{N}\right) & \frac{2}{N} & \frac{2}{N} & \cdots & \frac{2}{N} \\ \frac{2}{N} & \left(-1+\frac{2}{N}\right) & \frac{2}{N} & \cdots & \frac{2}{N} \\ \frac{2}{N} & \frac{2}{N} & \left(-1+\frac{2}{N}\right) & \cdots & \frac{2}{N} \\ \cdots & \cdots & \cdots & \cdots & \cdots \\ \frac{2}{N} & \frac{2}{N} & \frac{2}{N} & \cdots & \left(-1+\frac{2}{N}\right) \end{bmatrix}.$$

Note that as mentioned in the beginning of this section, the modulus of the sum of the entries in each column is indeed 1; hence it too may be viewed as a Markov Process with *transition probability* from any state to any other state being $\frac{2}{N}$. This fact will help in the design of the quantum search algorithm in the next section.

## 5. The Quantum Search Algorithm

It is easily seen that the maximum transfer into the marked state that can be accomplished by the matrix $D$ happens when the phase of the marked state is rotated by $\pi$ relative to the other states. This is illustrated in Figure 4, contrast this to the situation in section 3 where the maximum transfer was attained when the phase of the marked state was rotated by $\frac{\pi}{2}$.

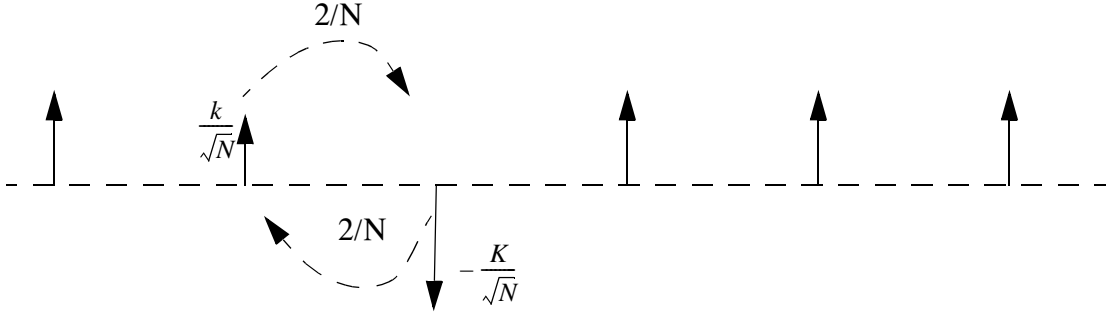

*Figure 4 - The diffusion transformation (D) results in a transfer of amplitude to the selected state whose phase is rotated by π by the R transformation. There is no net transfer between states with equal amplitude.*

Assume the amplitude in the marked state to be $-\frac{K}{\sqrt{N}}$, where $K \ll \sqrt{N}$, and the amplitude in each of the other states to be $\frac{k}{\sqrt{N}}$ where $k < 1$. As a result of $D$, as shown in Figure 4, there is a net transfer of $N \times \frac{2}{N}\left(\frac{k}{\sqrt{N}} + \frac{K}{\sqrt{N}}\right)$ into the marked state. When we add the initial amplitude of $-\frac{K}{\sqrt{N}}$, the total amplitude in the marked state after $D$ becomes $\frac{(K+2k)}{\sqrt{N}}$. Similarly, adding the initial amplitude to the amplitude transferred from the marked state, the amplitude in each of the other states becomes $\frac{k}{\sqrt{N}} - \frac{2}{N}\left(\frac{k}{\sqrt{N}}\right) - \frac{2K}{N\sqrt{N}}$ which is $\frac{k}{\sqrt{N}} - O\left(\frac{1}{N}\right)$. Therefore the net result of the operation is to change the amplitude in the marked state from $-\frac{K}{\sqrt{N}}$ to $\frac{(K+2k)}{\sqrt{N}}$ while leaving the amplitudes in the unmarked states approximately the same.

After this, if the amplitude in the marked state is inverted by $R$, the transformation $D$ can be used again to further increase the magnitude by $\frac{2k}{\sqrt{N}}$. As in section 3, $k$ stays virtually unaltered and is of order 1, it follows that in $O(\sqrt{N})$ repetitions of the $(DR)$ cycle, the amplitude in the marked state gets boosted to a quantity of order 1.

The algorithm can hence be described by the following sequence of transformations:

$$[\psi_{final}(\bar{x})] = \underbrace{(DR)............(DR)(DR)(DR)}_{\sqrt{N} \ reps}[\psi_{init}(\bar{x})]$$

Here $[\psi_{init}(\bar{x})]$ is the uniform superposition with equal amplitude of $\frac{1}{\sqrt{N}}$ in all $N$ states, $[\psi_{final}(\bar{x})]$ is a superposition which has an amplitude of order 1 in the marked state and small, uniform amplitudes in all other states. The matrices $D$ and $R$ are summarized below.

$$D \equiv \begin{bmatrix} \left(-1+\frac{2}{N}\right) & \frac{2}{N} & \frac{2}{N} & \cdots & \frac{2}{N} \\ \frac{2}{N} & \left(-1+\frac{2}{N}\right) & \frac{2}{N} & \cdots & \frac{2}{N} \\ \frac{2}{N} & \frac{2}{N} & \left(-1+\frac{2}{N}\right) & \cdots & \frac{2}{N} \\ \cdots & \cdots & \cdots & \cdots & \cdots \\ \frac{2}{N} & \frac{2}{N} & \frac{2}{N} & \cdots & \left(-1+\frac{2}{N}\right) \end{bmatrix} \quad \text{and} \quad R \equiv \begin{bmatrix} 1 & 0 & 0 & \cdots & 0 \\ 0 & -1 & 0 & \cdots & 0 \\ 0 & 0 & 1 & \cdots & 0 \\ \cdots & \cdots & \cdots & \cdots & \cdots \\ 0 & 0 & 0 & \cdots & 1 \end{bmatrix}$$

**6. Synthesizing $D$** The previous section described the matrix transformations which, when carried out, result in a significant amplitude being obtained in the marked state. These were $N \times N$ matrix transformations and we still need to show how to synthesize these by means of $O(\log N)$ two dimensional rotations and one dimensional reflections of the type discussed in section 1(ii) since these are the technologically feasible quantum computing operations.

By using the fact that $WW = I$ with $W$ as defined in section 1(ii), the transformation matrix $D$ as defined at the end of the previous section can be written in the following form: $W(WDW)W$. Substituting for $D$ from the previous section, it follows that:

$$D = -W\left(W\left(I - \frac{2}{N}\begin{bmatrix} 1 & 1 & 1 & \cdots & 1 \\ 1 & 1 & 1 & \cdots & 1 \\ 1 & 1 & 1 & \cdots & 1 \\ \cdots & \cdots & \cdots & \cdots & \cdots \\ 1 & 1 & 1 & \cdots & 1 \end{bmatrix}\right)W\right)W = -W\left(I - \frac{2}{N}W\begin{bmatrix} 1 & 1 & 1 & \cdots & 1 \\ 1 & 1 & 1 & \cdots & 1 \\ 1 & 1 & 1 & \cdots & 1 \\ \cdots & \cdots & \cdots & \cdots & \cdots \\ 1 & 1 & 1 & \cdots & 1 \end{bmatrix}W\right)W \qquad (a)$$

Assume that the states are encoded so that the first state is the one with all qubits in the 0 state (as mentioned before, this state is denoted by $\bar{0}$). Then since $M|0\rangle = \frac{1}{\sqrt{2}}(|0\rangle + |1\rangle)$, it follows that $W$ transforms $\bar{0}$ to a superposition with equal amplitudes in all $N$ states, i.e. $W\begin{bmatrix} 1 \\ 0 \\ 0 \\ \cdots \\ 0 \end{bmatrix} = \frac{1}{\sqrt{N}}\begin{bmatrix} 1 \\ 1 \\ 1 \\ \cdots \\ 1 \end{bmatrix}$. Since $WW = I$, it follows by multiplying both sides by $W$, that $W\begin{bmatrix} 1 \\ 1 \\ 1 \\ \cdots \\ 1 \end{bmatrix} = \sqrt{N}\begin{bmatrix} 1 \\ 0 \\ 0 \\ \cdots \\ 0 \end{bmatrix}$. Therefore:

$$W \begin{bmatrix} 1 & 1 & 1 & \dots & 1 \\ 1 & 1 & 1 & \dots & 1 \\ 1 & 1 & 1 & \dots & 1 \\ \dots & \dots & \dots & \dots & \dots \\ 1 & 1 & 1 & \dots & 1 \end{bmatrix} W = \sqrt{N} \begin{bmatrix} 1 & 1 & 1 & \dots & 1 \\ 0 & 0 & 0 & \dots & 0 \\ 0 & 0 & 0 & \dots & 0 \\ \dots & \dots & \dots & \dots & \dots \\ 0 & 0 & 0 & \dots & 0 \end{bmatrix} W = N \begin{bmatrix} 1 & 0 & 0 & \dots & 0 \\ 0 & 0 & 0 & \dots & 0 \\ 0 & 0 & 0 & \dots & 0 \\ \dots & \dots & \dots & \dots & \dots \\ 0 & 0 & 0 & \dots & 0 \end{bmatrix}.$$

(b)

Substituting (b) in (a):

$$D = -W \left( I - 2 \begin{bmatrix} 1 & 0 & 0 & \dots & 0 \\ 0 & 0 & 0 & \dots & 0 \\ 0 & 0 & 0 & \dots & 0 \\ \dots & \dots & \dots & \dots & \dots \\ 0 & 0 & 0 & \dots & 0 \end{bmatrix} \right) W = -W \begin{bmatrix} -1 & 0 & 0 & \dots & 0 \\ 0 & 1 & 0 & \dots & 0 \\ 0 & 0 & 1 & \dots & 0 \\ \dots & \dots & \dots & \dots & \dots \\ 0 & 0 & 0 & \dots & 1 \end{bmatrix} W$$

(c)

The equation (c) is written as $D = -W I_{\bar{0}} W$, where $I_{\bar{0}}$ denotes selective inversion of the $\bar{0}$ state.

## 7. Concluding remarks

This paper has outlined the steps that went into the development of the quantum search algorithm. As with many new scientific developments, the steps are far from rigorous. Nevertheless, I have found these insights helpful in developing the search algorithm and its extensions.

There have been several analyses of the search algorithm but there are still only partial answers to some basic questions which seems to suggest our understanding of this algorithm is still limited. The first question is: *What is the reason that one would expect that a quantum mechanical scheme could accomplish the search in $O(\sqrt{N})$ steps?* It would be insightful to have a simple two line argument for this without having to describe the details of the search algorithm.

It has been proved that the quantum search algorithm cannot be improved at all, i.e. any quantum mechanical will need at least $O(\sqrt{N})$ steps to carry out an exhaustive search of $N$ items [4] [5]. *Why is it not possible to search in fewer than $O(\sqrt{N})$ steps?* The arguments used to prove this are very subtle and mathematical. What is lacking is a simple and convincing two line argument that shows why one would expect this to be the case.

## 8. Appendix

- The quantum search algorithm is based on the assumption that it is possible to selectively invert the amplitudes in a basis state where the function $f(\bar{x})$ evaluates to 1. As mentioned in section 1(ii) such a circuit can be designed, provided we are given a quantum mechanical circuit to evaluate the function $f(\bar{x})$ in any specified basis state $\bar{x}$ (Figure 1). The selective inversion circuit is shown in Figure 5.

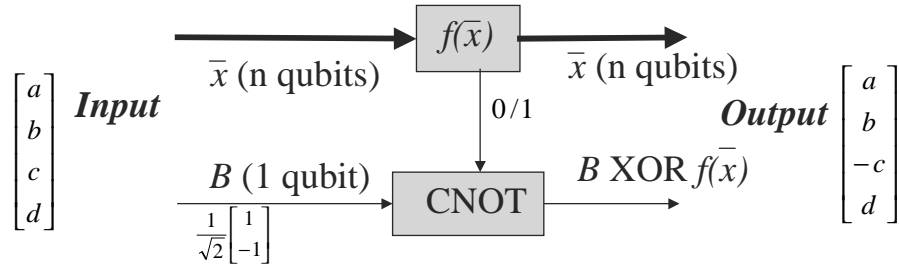

*Figure 5 - The above quantum mechanical circuit selectively inverts amplitudes of precisely those basis states where the function f(x) evaluates to 1.*

To analyze schematic quantum circuits, such as in Figure 5, one examines the transformation of the input basis states - and then by the superposition principle as described at the end of section 1(i), the effect on any superposition can be obtained. It is easily seen in the above circuit that if for some *n*-qubit input basis state $\bar{x}$, the output of the $f(\bar{x})$ gate is 1, the ancilla qubit superposition is transformed from $(\alpha, \beta)$ to $(\beta, \alpha)$. If $\alpha$ is $\frac{1}{\sqrt{2}}$ and $\beta$ is $-\frac{1}{\sqrt{2}}$, something very interesting happens. If the output of the $f(\bar{x})$ gate is 0, the ancilla bit superposition stays unchanged; whereas if the output of the $f(\bar{x})$ gate is 1, the ancilla qubit superposition is transformed from $\left(\frac{1}{\sqrt{2}}, -\frac{1}{\sqrt{2}}\right)$ into $\left(-\frac{1}{\sqrt{2}}, \frac{1}{\sqrt{2}}\right)$. This is equivalent to changing the sign of the amplitude. Thus in any *n*-qubit superposition, the amplitude of precisely those basis states are selectively inverted for which the function $f(\bar{x})$ is 1.